\documentclass[prl,showpacs,twocolumn]{revtex4}
\usepackage{graphics}
\usepackage{epsfig}
\usepackage{graphicx}

\begin{document}
\title{Entanglement in a Valence-Bond-Solid State}
\author{Heng Fan$^1$, Vladimir  Korepin$^2$, and Vwani Roychowdhury$^1$} 
\affiliation{$^1$Electrical Engineering Department,
University of California at Los Angeles, Los Angeles,
CA 90095, USA\\
$^2$C.N.Yang Institute for Theoretical Physics,
State University of New York at Stony Brook, Stony Brook,
NY 11794-3840, USA}

\date{\today}

\begin{abstract}
We study entanglement in Valence-Bond-Solid
state, which describes the ground state of an AKLT quantum spin chain,
consisting of  bulk spin-1's and
two spin-1/2's at the ends. We characterize entanglement between various
subsystems of the ground state by mostly calculating the entropy of one
of the subsystems;  when appropriate, we evaluate concurrences as
well. We show that the reduced density matrix of a continuous block of
bulk spins is independent of the size  of the chain and the
location of the block relative to the ends. Moreover, we show that the
entanglement of the block with the rest of the sites approaches a
constant value exponentially fast, as the size of the block increases. 
We also calculate the  entanglement of (i) any two bulk spins 
with the rest, and (ii)  the end
spin-1/2's (together and separately) with the rest of the ground state.
For example, we
show that (i) any two bulk spins become maximally entangled with the
rest of the ground state  exponentially fast in their separation
distance, (ii) the two end
spin-1/2's share no entanglement, and (iii) each end spin-1/2 is
maximally entangled with the rest.
\end{abstract}

\maketitle

There is  considerable current interest in quantifying entanglement
in various quantum systems.
Entanglement in spin chains, correlated electrons, interacting bosons
and other  models was studied  
\cite{OAFF,ON,V,VLRK,LRV,JK,K,ABV,VMC,LO,PP,FS,V2,keat,xy,salerno, zanardi,
rasetti, cardy,honk1,honk2, kais1, kais2,hlw}.
Entanglement is a fundamental measure of how much quantum effects 
we can observe and use, and it is the primary resource in
quantum computation and quantum information processing \cite{BD,L}.
Also entanglement  plays a
role in the quantum phase transitions \cite{OAFF,ON}, and it has been
experimentally demonstrated that the entanglement
may affect macroscopic properties of solids \cite{GRAC,V}.

In this Letter, we will study a spin chain introduced by Affleck,
Kennedy, Lieb, and Tasaki (AKLT model) \cite{AKLT,AKLT0}.
The ground state of the model  is a unique pure state. It is known as
Valence-Bond Solid (VBS), and plays a central role in condensed matter
physics. Haldane \cite{H} conjectured that 
an anti-ferromagnetic  Hamiltonian describing  half-odd-integer spins
is gap-less, but for integer spins it has a gap. AKLT model describing
interaction of  spin-1's in the bulk agrees with the  conjecture.  
An implementation of AKLT in optical lattices was proposed recently \cite{GMC}, 
and the use of AKLT 
model  for  {\it universal 
quantum computation} was discussed in \cite{VC}. VBS is also closely related 
 to Laughlin ansatz \cite{L0} and 
to fractional quantum Hall effect \cite{AAH}.

We investigate the seminal AKLT model from the 
new perspective of quantum information, 
and 
evaluate the entanglement (in terms of entropy)  
of various subsystems of the VBS.
The results and the methodologies adopted herein have several implications
from the perspective of both quantum information and condensed matter. For 
example, 
while the entanglement in spin chains with periodic boundary conditions
has been studied extensively, our results provide entanglement calculations
for spin chains with  open boundary conditions.
For critical gap-less models conformal field theory describes the entanglement
 \cite{hlw,K,cardy}. 
 For gapped models  G. Vidal,  J.I.Latorre, E.Rico and A.Kitaev \cite{VLRK} 
 conjectured that the entropy of a  large  block of spins reach saturation.
We confirm this for  the AKLT model, and find that the entropy of a large block of
bulk spins is close to two. This means that the block can be in four
different states, and hence, the Hilbert space of states of the large block
of bulk spins is four-dimensional.
Our results also show that the entanglement correlation of VBS state is
{\it short-ranged}, which provides a good understanding why the
density matrix renormalization group (DMRG) method \cite{W0}
works so efficiently for VBS states; see \cite{VGC} for recent developments.

The AKLT model consists of a linear 
chain of $N$ spin-1's in the bulk,  
and two spin-1/2's  on the boundary. 
We shall denote by $\vec{S}_k$  the vector of
spin-1 operators and 
by  $\vec{s}_b$ spin-1/2 operators, where $b=0,N+1$.
The Hamiltonian is:
\begin{eqnarray} 
H=\sum _{k=1}^{N-1}
\left( \vec {S}_k 
\vec {S}_{k+1}
+{1\over 3}(\vec {S}_k 
\vec {S}_{k+1})^2
\right)+ \pi_{0,1} +\pi_{N, N+1} .
\end{eqnarray} 
The boundary terms $\pi $ describe interaction of a spin 1/2 and spin 1.
Each term is a projector on a state with  spin 3/2:
\begin{eqnarray}
\pi_{0,1}={\frac{2}{3}}\left(1+\vec{s}_0\vec{S}_1  \right),\quad 
\pi_{N, N+1}= {\frac{2}{3}}\left(1+\vec{s}_{N+1}\vec{S}_N  \right) .
\end{eqnarray}

The ground state of this model is unique and can be represented  as
\cite{AKLT,AKLT0}:
\begin{eqnarray}
|G\rangle (\otimes _{k=1}^NP_{k\bar {k}})
|\Psi ^-\rangle _{\bar {0}1}
|\Psi ^-\rangle _{\bar {1}2}\cdots 
|\Psi ^-\rangle _{\bar {N}N+1}. 
\label{ground1}
\end{eqnarray}
Here $P$  projects
a state of two qubits on a symmetric subspace, which describes spin $1$.
In the formula above  $|\Psi ^-\rangle (|\uparrow \downarrow \rangle -
|\downarrow \uparrow \rangle )/\sqrt{2}$ represents a singlet state, 
and the subscripts represent the two parties the 
singlet is shared between.
We have tried to keep our notations as close to 
those in the paper \cite{VMC}. We can use the 
following figure to visualize the
ground state:
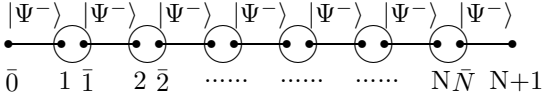
\begin{center}
\unitlength=1mm
\begin{picture}(80,5)(0,3)
\put(2,0){\makebox(3,3){$|\Psi ^-\rangle $}}
\put(12,0){\makebox(3,3){$|\Psi ^-\rangle $}}
\put(22,0){\makebox(3,3){$|\Psi ^-\rangle $}}
\put(32,0){\makebox(3,3){$|\Psi ^-\rangle $}}
\put(42,0){\makebox(3,3){$|\Psi ^-\rangle $}}
\put(52,0){\makebox(3,3){$|\Psi ^-\rangle $}}
\put(62,0){\makebox(3,3){$|\Psi ^-\rangle $}}
\end{picture}
\begin{picture}(80,5)(0,0)
\put(0,0){\line(1,0){7}}
\put(10,0){\line(1,0){7}}
\put(20,0){\line(1,0){7}}
\put(30,0){\line(1,0){7}}
\put(40,0){\line(1,0){7}}
\put(50,0){\line(1,0){7}}
\put(60,0){\line(1,0){7}}
\put(0,0){\circle*{1}}
\put(7,0){\circle*{1}}
\put(10,0){\circle*{1}}
\put(17,0){\circle*{1}}
\put(20,0){\circle*{1}}
\put(27,0){\circle*{1}}
\put(30,0){\circle*{1}}
\put(37,0){\circle*{1}}
\put(40,0){\circle*{1}}
\put(47,0){\circle*{1}}
\put(50,0){\circle*{1}}
\put(57,0){\circle*{1}}
\put(60,0){\circle*{1}}
\put(67,0){\circle*{1}}
\put(8.5,0){\circle{5}}
\put(18.5,0){\circle{5}}
\put(28.5,0){\circle{5}}
\put(38.5,0){\circle{5}}
\put(48.5,0){\circle{5}}
\put(58.5,0){\circle{5}}
\end{picture}
\begin{picture}(80,5)(1,1)
\put(0,0){\makebox(3,3){$\bar {0}$}}
\put(7,0){\makebox(3,3){1}}
\put(10,0){\makebox(3,3){$\bar {1}$}}
\put(17,0){\makebox(3,3){2}}
\put(20,0){\makebox(3,3){$\bar {2}$}}
\put(27,0){\makebox(3,3){...}}
\put(30,0){\makebox(3,3){...}}
\put(37,0){\makebox(3,3){...}}
\put(40,0){\makebox(3,3){...}}
\put(47,0){\makebox(3,3){...}}
\put(50,0){\makebox(3,3){...}}
\put(57,0){\makebox(3,3){N}}
\put(60,0){\makebox(3,3){$\bar {N}$}}
\put(67,0){\makebox(3,3){N+1}}
\end{picture}
\end{center}
\medskip
A black dot represents  spin-$\frac {1}{2}$, and 
spin-$1$'s are denoted by  circles. To begin with, 
each bulk site, $k$ (where
$1\leq k\leq N$) shares one singlet 
state $|\Psi ^-\rangle $ (represented by a line) 
 with its left and right neighbors. Thus
at each bulk site, $k$,  we  start with two spin-$1/2$'s labeled by 
$(k,\bar {k})$  and then the spin-1's are 
prepared by projecting the two spin-1/2's (4-dimensional space)  
on a symmetric three  dimensional subspace of spin $1$ (3-dimensional). 
The system has  open boundary conditions, and 
the two ends are numbered as sites $\bar {0}$ (before projection, 
this site shared a singlet with site $1$) 
and  $N+1$.

There is an upper bound on the entropy of a block of
$L$ spins. Before projection, the entropy is equal
to 2,  since  the boundary intersects two singlet states. 
Since the local projections will only decrease the
entanglement, we expect that the entropy of a block
of $L$ spins to have an upper bound of $2$. 

In order to calculate the reduced density 
matrices of various subsystems of the 
ground state $|G\rangle$ (see Eq.~\ref{ground1}), 
it is more convenient to express it in a different 
form based on the singlet chain shown in the 
preceding figure and the figure below.
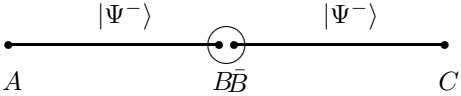
\begin{center}
\unitlength=1mm
\begin{picture}(80,5)(0,3)
\put(14,0){\makebox(3,3){$|\Psi ^-\rangle $}}
\put(44,0){\makebox(3,3){$|\Psi ^-\rangle $}}
\end{picture}
\begin{picture}(80,5)(0,0)
\put(0,0){\line(1,0){28}}
\put(30,0){\line(1,0){28}}
\put(0,0){\circle*{1}}
\put(28,0){\circle*{1}}
\put(30,0){\circle*{1}}
\put(58,0){\circle*{1}}
\put(29,0){\circle{5}}
\end{picture}
\begin{picture}(80,5)(1,1)
\put(0,0){\makebox(3,3){$A$}}
\put(28,0){\makebox(3,3){$B$}}
\put(30,0){\makebox(3,3){$\bar {B}$}}
\put(58,0){\makebox(3,3){$C$}}
\end{picture}
\end{center}
\medskip
Let us first consider a chain of
two singlet states, $|\Psi ^-\rangle _{AB}$ 
and $|\Psi ^-\rangle _{\bar{B}C}$:
$A$ is in
site \#1, ($B$, $\bar {B}$) is in site \#2, and $C$ is
in site \#3. The combined state can then be expressed as follows:
\begin{eqnarray}
&&|\Psi ^-\rangle _{AB}|\Psi ^-\rangle _{\bar{B}C}
={1\over 2}\sum _{\alpha =0}^3\left( 
(-)^{1+\alpha }I_B\otimes (\sigma _{\alpha }^*)_{\bar {B}}
\right.
\nonumber \\
&&\left. \otimes I_A\otimes (\sigma _{\alpha })_C\right)
|\Psi ^-\rangle _{B\bar {B}}|\Psi ^-\rangle _{AC},
\label{swap}
\end{eqnarray} 
where both $I$ and $\sigma _0$ represent the identity operator, 
$\sigma _1,\sigma _2,\sigma _3$ are the Pauli 
matrices, and `*' means  complex
conjugation. 
By entanglement swapping similar to teleportation \cite{BBC},
party \#2 can perform a Bell state 
measurement on ($B$, $\bar{B}$), and then
communicate the results of measurements to  party \#1 or \#3.
Then one of them can perform a  unitary transformation locally, and finally
a maximally entangled state will be shared by them. 
A multi-dimensional generalization of this  can be found, 
for example in \cite{F}.   

Eq.(\ref{swap}) can  be generalized to a chain of 
singlet states. First, define quantum states
$|\alpha \rangle =(-1)^{1+\alpha }
(I\otimes \sigma _{\alpha }^*)|\Psi ^-\rangle $. 
Thus, $|0\rangle $ is the
singlet state with spin 0, while other three states $|1\rangle ,
|2\rangle ,|3\rangle $ form the symmetric 
subspace of spin-1 (within a phase).
Repeatedly using
the relation (\ref{swap}) and with the help of
the property presented later
in the proof of our theorem, we obtain:
\begin{eqnarray}
&&|\Psi ^-\rangle _{\bar {0}1}
|\Psi ^-\rangle _{\bar {1}2}\cdots 
|\Psi ^-\rangle _{\bar {N}N+1}
=\frac {1}{2^N}\sum _{\alpha _1,\cdots ,\alpha _N=0}^3
|\alpha _1\rangle \cdots 
\nonumber \\
&&~~~\cdots 
|\alpha _N\rangle 
\left( I_{\bar {0}}\otimes (\sigma _{\alpha _N}\cdots 
\sigma _{\alpha _1})_{N+1}
\right)
|\Psi ^-\rangle _{\bar {0},N+1}.
\end{eqnarray}
The quantum states $|\alpha _i\rangle $ are orthonormal states
at lattice site $(i,\bar {i})$. 
Thus, by projecting the quantum state on the symmetric 
subspace spanned by the states $|1\rangle$, $|2\rangle$, 
and $|3\rangle$, the ground state of AKLT model can be 
rewritten as \cite{VMC,FNW}:
\begin{eqnarray}
|G\rangle 
&&=\frac {1}{3^{N/2}}\sum _{\alpha _1,\cdots ,\alpha _N =1}^3
|\alpha _1\rangle \cdots 
\nonumber \\
&&\cdots |\alpha _N\rangle 
(I_{\bar {0}}\otimes (\sigma _{\alpha _N}\cdots \sigma _{\alpha _1})_{N+1})
|\Psi ^-\rangle _{\bar {0},N+1}.
\label{ground2}
\end{eqnarray}
It follows directly from Eq.(\ref{ground2}) that the 
reduced density matrix of spin-1 at any bulk site $k$ 
(recall that  $k=1,...,N$) is:
\begin{eqnarray}
\rho _1\equiv {\rm Tr}_{1,...\{ k\}...,N,\bar {0},N+1}|G\rangle \langle G|
=\frac {1}{3}\sum _{\alpha _k=1
}^3|\alpha _k
\rangle \langle \alpha _k|, 
\label{onesite}
\end{eqnarray}
where the trace is taken over all sites (including the two ends), 
except site number $k$. We see that all {\it 
one-site reduced density operators in the
bulk are the same}:  
the identity or the maximally-disordered state in the spin-1 space.  
Thus, the single-site reduced density matrices are  
independent of the total size of the 
spin chain $N$, and of the distance from the
ends (i.e., $k$ or $N-k$).  
For the more general case, we have the following result:

\noindent
{\bf Theorem}: {\it Consider the reduced density matrix of a 
continuous block of spins of  length $L$ (not 
including the two boundary 1/2-spins), 
starting from site $k$ and stretching up 
to $k+L-1$, where $k \geq 1$ 
and $k+L-1\leq N$ (thus, $1\le L \le N$) in the VBS  ground state 
(\ref{ground2}). Then, all these  
density operators are the same, and independent of both 
$k$ (i.e., the location of the block) 
and of $N$ (the total length of the chain). 
Thus, the reduced density matrix depends only on $L$, 
the length of the block under consideration.}

The proof is based  on the following relations:
Define $|\Phi ^+\rangle =(|\uparrow \uparrow \rangle
+|\downarrow \downarrow \rangle )/\sqrt{2}$, 
we know that $|\Phi ^+\rangle =(-i)(\sigma _2\otimes I)|\Psi ^-\rangle $.
For a unitary operator $U$, we have the property 
$(U\otimes U^*)|\Phi ^+\rangle =|\Phi ^+\rangle $.
Then
$(U_1\otimes U_2)|\Phi ^+\rangle =(U_1U_2^t\otimes I))|\Phi ^+\rangle 
=(I\otimes U_2U_1^t)|\Phi ^+\rangle $, where $U_1,U_2$ are two
unitary operators (the super-index $t$ denotes the transposition).

By using these relations, we can prove that:
\begin{eqnarray}
&&{\rm Tr}_{\bar {0},N+1}(I\otimes U_1VU_2)
|\Psi ^-\rangle \langle \Psi ^-|
(I\otimes U_1V'U_2)^{\dagger }
\nonumber \\
&=&{\rm Tr}_{\bar {0},N+1}(I\otimes V)|\Phi ^+\rangle \langle \Phi ^+|
(I\otimes V')^{\dagger }.
\end{eqnarray}
By repeated applications of this relation, and considering the ground
state  (\ref{ground2}), the reduced density operator of any
continuous block of spins of length $L$ is
\begin{eqnarray}
\rho _L&=&\frac {1}{3^L}\sum _{\alpha ,\alpha '}
|\alpha _1\rangle \langle \alpha _1'|
\cdots |\alpha _L\rangle \langle \alpha _L'|\times
\nonumber \\
&&{\rm Tr}_{\bar {0},N+1}(I\otimes V)|\Phi ^+\rangle \langle \Phi ^+|
(I\otimes V')^{\dagger }.
\end{eqnarray}
where $V=\sigma _{\alpha _L}\cdots \sigma _{\alpha _1},
V'=\sigma _{\alpha _L'}\cdots \sigma _{\alpha _1'}$.
This operator only depends on $L$. This completes our proof.

Our aim is to calculate the {\it entanglement} of the VBS state.
For a pure bi-partite state $|\psi \rangle _{AB}$, the
entanglement between spatially separated parties $A$ and $B$
is $S(\rho _A)=S(\rho _B)$, where $\rho _{A(B)}{\rm Tr}_{B(A)}|\psi \rangle \langle \psi |$ are the reduced
density operators and 
$S(\rho )=-{\rm Tr}\rho \log \rho $
is the von Neumann {\it entropy}, where we take the 
logarithms in the base $2$. For example, 
it follows from Eq.~(\ref{onesite}) that the entropy
of the one-site reduced density operator in the bulk is
$S\left( \rho _1(k)\right) =\log 3$.
This entropy describes the entanglement between site number $k$ in
the bulk (considered as one party) and the rest of 
the ground state (considered as the other party). 
The space of spin-1 is three dimensional, 
so $\log 3$ is the maximum of the entropy.
So we proved  that in the VBS state (\ref{ground2}), 
each individual spin in the bulk is maximally entangled 
with the rest of the ground state. Later in the paper,  
we shall see that this is also true for the boundary spin-1/2's.

Since the reduced density operator   of  a  continuous block
of $L$ spins is  independent of the total  size, $N$, 
of the spin chain, we can consider the case where
$L=N$, i.e., we consider a chain of $L$ spin-1's with 
one spin-1/2 at each end.
Now the reduced density operator of two end spin-1/2's
takes the following form:
\begin{eqnarray}
\rho _{\hat {L}}&=&\frac {1}{3^L}\sum _{\alpha _1,\cdots ,\alpha _L=1}^3
(I\otimes \sigma _{\alpha _L}
\cdots \sigma _{\alpha _1})|\Psi ^-\rangle
\langle \Psi ^-| \times
\nonumber \\
&&\times (I\otimes \sigma _{\alpha _L}\cdots \sigma _{\alpha 1})^{\dagger }= 
\nonumber \\
&=&\frac {1}{4}
\left( 1-p(L)\right) \cdot I
+p(L)|\Psi ^-\rangle
\langle \Psi ^-|.
\label{twoends} 
\end{eqnarray}
Here  $p(L)=(-1/3)^L$ and $I$ is the identity in 4 dimensions.
Since the ground state (\ref{ground2}) is  
pure, the entropy of the block of $L$ bulk spin-1's   is equal to the
entropy of  the two ends.   So we have
\begin{eqnarray}
S_L&\equiv & S(\rho _L)=S(\rho _{\hat {L}})= 
\nonumber  \\ \nonumber
&=&2+ \frac {3\left( 1-p(L)\right) }{4}\log \left( 1-p(L)\right) - 
\\ \label{entalpy}
&-&\frac {1+3p(L)}{4}\log \left( 1+3p(L)\right). 
\end{eqnarray}
As expected, $S_L \leq 2$ and approaches two $2$ exponentially fast in $L$:
$S_L\sim 2 -(3/2)p(L)$. 
This is also clear from (\ref{twoends}): 
the reduced density operator 
approaches the identity in the 4-dimensions exponentially fast.
Consider the numbers:
\begin{eqnarray}
\begin{array}{lll}
S_1=1.58496 &S_2=1.97494 & S_3=1.99695
\\
S_4=1.99969 & S_5=1.99996 &S_6\approx 2.
\end{array}
\end{eqnarray} 
Note that the correlation function of local spins   decays equally fast: 
\begin{eqnarray}
<\vec {S}_L \vec {S}_{1}>\sim  (-1/3)^{L}=p(L), \label{lc}
\end{eqnarray}
see  \cite{AKLT0,AAH}.

Next we shall study the entropy of  {\bf two spin-1}'s 
separated by $M$ sites in the bulk. That is we calculate 
the entanglement between two two bulk spin-1's and 
the rest of the spin-1's and the two spin-1/2's. 
We still can  show that the reduced density operator  does not
depend on the total size of the  chain, $N$, and prove that:
\begin{eqnarray}
\rho _2(M)=\frac {1}{9}(1-p(M))I+p(M)\rho _2,  
\end{eqnarray}
where $p(M)=(-1/3)^M$ and  $\rho _2$ is the two-site
reduced density operator of nearest neighbors, i.e. the case  $M=0$, and
the operator $I$  is the identity  in nine-dimensions.
The nearest neighbor two-site reduced density
operator can be written explicitly:
\begin{eqnarray}
&&\rho _2=\frac {1}{9}[\sum _{\alpha ,\beta =1}^3
|\alpha \rangle \langle \beta |\otimes |\alpha \rangle
\langle \beta | +
\nonumber \\
&&+
\sum _{\alpha \not= \beta }(|\alpha \rangle 
\langle \alpha |\otimes |\beta \rangle \langle \beta |
-|\alpha \rangle \langle \beta |\otimes |\beta \rangle
\langle \alpha |)] .
\end{eqnarray}   
So we can calculate the entropy of two spins at  distance $M$:
\begin{eqnarray}
&&S_2(M)=2\log 3-\frac {5}{9}(1-p(M))\log (1-p(M))-
\nonumber \\
&&-\frac {3}{9}(1+p(M))\log (1+p(M)) -\nonumber \\
&&-\frac {1}{9}(1+2p(M))\log (1+2p(M)). \label{two-spins}
\end{eqnarray}
We see that   $S_2(M)$  also  approaches the maximum value (since the
dimension is $9$, the maximum entropy is $2\log 3$)
 with the exponential  rate defined by local correlations (\ref{lc}). 
Note that $S_2=S_2(0)$ (see Eq.(\ref{entalpy})) 
and (\ref{two-spins})). However, for $M\geq 1$, $S_2(M)$ 
quickly exceeds $S_L$.
We  also can calculate the {\it concurrence}  
(another   measure of entanglement
\cite{W}). We shall use the generalized concurrence 
in higher dimensions \cite{FMH}.
Two concurrences corresponding to $S_L$ and $S_2(M)$ are equal to
$\displaystyle
C_L= 1-p^2(L)=1-\frac {1}{9^L}$ and $\displaystyle 
C_2(M) = 1-\frac {1}{6}{p^2(M)}=1-\frac {1}{6\cdot 9^M}$, respectively. 
They look similar because the entanglement of the block  also
represents the entanglement of two ends with $L$ bulk spins.

Now we  turn to the analysis of entanglement of {\bf boundary} spins.
We  start from the reduced density operator of one boundary spin.
We can prove that  it is the identity matrix in two-dimensions.
This shows  that the end spin-1/2's  are maximally entangled 
with the rest of
the ground state, and has an entropy of  1.

The density operator of two ends $\rho _{\hat {N}}$ 
(see Eq.(\ref{twoends})) depends on
the total size of the lattice  $N$. In  the simplest case,
$\rho _{\hat 1}=(I-|\Psi ^-\rangle \langle \Psi ^-|)/3$.
This is a  separable state. Actually it is separable for any $N$.
So, there is no entanglement  between the two ends. 
If the size of the spin chain $N$ increases, 
$\rho _{\hat {N}}$   approaches quickly the 
identity matrix in four dimensions.
In Eq.(~\ref{entalpy}), 
replacing $L$ by $N$ in $S_L$ we get the  
entanglement  between the two ends (one subsystem) and all
$N$ bulk  spins (another subsystem). 
It means that two ends considered as a subsystem, 
are maximally entangled with the bulk
spins if $N$ is large.  

Next we consider two-site reduced density operator
with {\bf one spin in the bulk} and {\bf another  spin at an one end}.
It is enough to put  the end spin at site  $\bar {0}$,
and the bulk spin at the site  $(M+1)$ (the range is $M=0,...,N-1$).
We can calculate the reduced density operator as above:
\begin{eqnarray}
&&\rho _2(\bar {0},M+1)
=\frac {1}{6}I
+\frac {p(M)}{6} 
[|1\rangle \langle 2|\otimes i(|0\rangle \langle 0|
-|1\rangle \langle 1|)
\nonumber \\
&&+|2\rangle \langle 1|\otimes i(|1\rangle \langle 1|
-|0\rangle \langle 0|)
+|1\rangle \langle 3|\otimes (|1\rangle \langle 0|
-|0\rangle \langle 1|)
\nonumber \\
&&+|3\rangle \langle 1|\otimes (|0\rangle \langle 1|
-|1\rangle \langle 0|)
+|2\rangle \langle 3|\otimes i(|0\rangle \langle 1|
+|1\rangle \langle 0|)
\nonumber \\
&&+|3\rangle \langle 2|\otimes (-i)(|0\rangle \langle 1|
+|1\rangle \langle 0|)],
\end{eqnarray}
where $p(M)=(-1/3)^M$.
First we consider if this state is separable.
Since it is the $2\times 3$-dimensional case,
we can use Peres-Horodecki  criterion \cite{P,H1}.
We find that when $M=0$, the state is  entangled. For
$M\not =0$  it is a separable state.
So, we know that the end spin $\bar {0}$  is  entangled only with
its nearest neighbor (spin-1). Secondly, we can  study the entropy
of this  state,
it is:
\begin{eqnarray}
&&S(\rho _2(\bar {0},M+1))=\log 6-
\frac {2}{3}(1-p(M))\log (1-p(M))
\nonumber \\
&&-\frac {1}{3}(1+2p(M))\log (1+2p(M)).
\end{eqnarray}
Similar to other entropies presented above, 
it approaches the upper bound
 $\log 6$ with the same exponential speed, defined by local correlations
 (\ref{lc}).
The concurrence corresponding to this
entanglement is: $C(\bar {0},M+1)=1-\frac {2}{5}{p^2(M)}$.

In summary, 
the entanglement properties of the VBS state 
can be listed as follows: ({\bf 1}) each
individual spin is maximally entangled with the rest; ({\bf 2}) the entanglement of a block of spins of length $L$ with the rest gets to
a constant value exponentially fast with $L$; ({\bf 3}) the
entanglement of any two bulk spins gets maximal exponentially fast in
their distance; ({\bf 4}) each individual boundary spin is
maximally entangled with its nearest neighbor and not with the other
bulk spins and the other boundary spin; and ({\bf 5}) each individual
boundary spin and another individual bulk spin are entangled with the
rest, and the entanglement gets maximal exponentially fast with the distance
between the boundary spin and its bulk partner. 

In the future it will be
interesting to calculate the entropy of a subsystem of
2 and 3-dimensional AKLT model; in fact, we are planning
to study the entanglement for AKLT on arbitrary graphs
using the results of  \cite{KK}. We believe that it will be
 useful for universal quantum computation, as in \cite{KLO}.

Acknowledgments: We are grateful to Doctor Bai Qi Jin for discussions. 

\end{document}